\documentclass{ws-p9-75x6-50}

\usepackage{graphics}
\usepackage{psfig}

\begin{document}

\title{ The iron $K_\alpha$-line diagnostics of
a rotational black hole metric
}

\author{Alexander F. Zakharov}
\address{Institute of Theoretical and Experimental Physics, Moscow
 117259,  Russia;\\
Astro Space Centre of Lebedev Physics Institute, Moscow, Russia,\\
%Dipartimento di Fisica Universita di Lecce and INFN, Sezione di
%Lecce,  Italy \\
 E-mail: zakharov@vitep1.itep.ru}
\author{ Serge V. Repin}
\address{Space Research Institute, %\\
%84/32, Profsoyuznaya st.,
             Moscow, 117810, Russia.\\
           E-mail: repin@mx.iki.rssi.ru}


\maketitle

\abstracts{ The original idea to show the spacetime geometry
using few geodesics was developed by Johnson and Ruffini (1974).
We used this idea to interpret the observational data for rotating
BH's. We developed the imitation approach to simulate a
propagation of radiation near BH's. An important problem for this
approach is the diagnostics of a black hole metric using X-ray
observational data of the iron $K_\alpha$-line.
    Observations of Seyfert galaxies in X-ray region reveal the
broad emissiion lines in their spectra, which can arise in inner
parts of accretion disks, where the effects of General Relativity
(GR) must be counted. A spectrum of a solitary emission line (the
$K_\alpha$-line of iron, for example) of a hot spot in Kerr
accretion disk is simulated, depending on the radial coordinate
$r$ and the angular momentum $a=J/M$ of a black hole, under the
assumption of an equatorial circular motion of a hot spot. Using
results of numerical simulations it is shown that the
characteristic two-peak line profile with the sharp edges arises
at a large distance, (about $r \approx (3-10)r_g$). The inner
regions emit the line, which is observed with one maximum and
extremely broad red wing. High accuracy future spectral
observations, being carried out, could detect the angular
momentum $a$ of the black hole. We analyzed the different
parameters of problems on the observable shape of this line and
discussed some possible kinds of these shapes. The total number
of  geodesics is about $10^9$ (to simulate possible shapes of the
$K_\alpha$-line), so the number is great enough, especially in
comparison with few geodesics in the original paper by Johnson
and Ruffini (1974). }

It is clear that an analysis of geodesics gives the direct way to
investigate a metric. In particular, Johnson and
Ruffini\cite{John74} calculated geodesics near a Kerr black hole.
These geodesics were very beautiful and they were used to  create
the known ICRA emblem. However, there is a problem to recognize
such shapes of these geodesics near black holes, because
practically we use something like photon geodesics to look at the
geodesics near black holes, but the photon geodesics are bent by
a very complicated way, so it is extremely difficult to
reconstruct the shapes of geodesics near black holes. Below we
will discuss how to extract an information about a metric from an
analysis of geodesics around black holes.

The general status of black holes described in a number of papers
(see, for example the following
papers\cite{Liang98,Zak00,FN01} and references therein).
As it was emphasized in these reviews the most solid evidence for
an existence of black holes comes from observations of some Seyfert
galaxies because we need a strong gravitational field approximation
to interpret these observational data, so probably we observe
manifestations radiation processes from the vicinity of the
black hole horizon (these regions are located inside the
Schwarzschild black hole horizon, but outside the Kerr black hole
horizon, thus we should conclude that we have manifestations
of rotational black holes).

    Recent observations of Seyfert galaxies in X-ray
    band\cite{fabian1,tanaka1,nandra1,nandra2,malizia,sambruna,fabian2}
reveal
the existence of wide iron $K_\alpha$ line (6.4~keV) in their
spectra along with a number of other weaker lines
(Ne~X, Si~XIII,XIV, S~XIV-XVI, Ar~XVII,XVIII, Ca~XIX,
etc).\cite{fabian1,tanaka1,nandra1,nandra2,malizia,sambruna,fabian2}
The line width corresponds to the velocity of the matter motion
of tens of thousands kilometers per second, reaching the maximum
value $v \approx 80000 - 100000$~km/s for the galaxy
MCG-6-30-15\cite{tanaka1} and $v \approx 48000$~km/s
for MCG-5-23-16.\cite{krolik1}
In some cases the line has characteristic two-peak
profile
with a high ``blue'' maximum and
the low ``red'' one and the long red wing, which gradually
drops to the background level.\cite{tanaka1,yaqoob2}

To simulate these shapes of the spectral lines we choose
a minimal number of assumptions.
We used the numerical approach based on the method,
described earlier.\cite{zakh91,zakharov6,zakharov1,zakharov5,zak_rep1,zak_rep2}

%\section{Geodesics in a Kerr metric}

Many astrophysical processes,
%with great observed energy release
where the great energy release is observed,
are assumed to be connected with the black holes.
Because the main part of the astronomical objects, such as the stars
and galaxies, possesses the proper rotation, then there are no doubts
that the black holes, both stellar and supermassive, possess the
intrinsic proper rotation too. Therefore we consider an emission of
monoenergetic quanta near a Kerr black hole.

     The large amount of observational data requires its
comprehension, theoretical simulation and interpretation. The
numerical simulations of the accretion disk spectrum under GR
assumptions has been reported in the paper.\cite{bromley} The
observational manifestations of GR effects are considered in
X-ray binaries\cite{cui}. Different physical models of the origin
of a broad emission iron $K_\alpha$ line in the nuclei of Seyfert
galaxies are analyzed in the papers.\cite{fabian2,sulentic2}

     The stationary black holes are described by the Kerr
metric:\cite{wheeler}
\begin{eqnarray}
  ds^2 = %&
\displaystyle
          - \frac{\Delta}{\rho^2}
            \left(dt - a\sin^2\theta d\phi\right)^2 +
          % \hspace{33mm}                              \nonumber\\
         %&
\displaystyle
           \frac{\sin^2\theta}{\rho^2}
              \left[\left(r^2 + a^2\right) d\phi - a dt \right]^2 +
            \frac{\rho^2}{\Delta} dr^2 + \rho^2 d\theta^2,
                           \label{eq1}
\end{eqnarray}
where
\begin{equation}
   \rho^2 = r^2 + a^2 \cos^2\theta,
\quad
\end{equation}
\begin{equation}
   \Delta = r^2 - 2Mr + a^2.
\end{equation}

     The photons trajectories can be described by the standard
 equations of geodesics:
\begin{equation}
\frac{d^2 x^i}{d\lambda^2}+\Gamma^i_{kl}
\frac{dx^k}{d\lambda}
\frac{dx^l}{d\lambda}=0,
 \label{eq4}
\end{equation}
 where  $\Gamma^i_{kl}$ are the Christoffel symbols.
The equations geodesics however can be simplified if we will use
the complete set of the first integrals which were found by Carter:\cite{carter}
$E=p_t$ is the particle energy at infinity, $L_z = p_\phi$ is $z$-component
of its angular momentum, $m=p_ip^i$ is the particle mass and
$Q$ is the Carter's separation constant:
\begin{equation}
   Q = p_\theta^2 + \cos^2\theta
                    \left[a^2 \left(m^2 - E^2\right) +
                         %\frac
{L_z^2}/{\sin^2\theta}
                    \right].     \label{eq5}
\end{equation}
As shown by Zakharov,\cite{zakh91,zakharov1}
the equations of photon motion can be reduced to
\begin{eqnarray}
   \frac{dt^\prime}{d\sigma}
                           & = &
      - a \left(a \sin^2\theta - \xi\right) +
      \frac{r^2 + a^2}{\Delta}
       \left(r^2 + a^2 - \xi a\right),
                        \label{eq6}                       \\
   \frac{dr}{d\sigma} & = & r_1,       \label{eq7}        \\
   \frac{dr_1}{d\sigma}    & = &
      2r^3 + \left(a^2 - \xi^2 - \eta\right) r +
      \left(a - \xi\right) + \eta,                        \\
   \frac{d\theta}{d\sigma} & = & \theta_1,                \\
   \frac{d\theta_1}{d\sigma}
                           & = &
      \cos\theta \left(\frac{\xi^2}{\sin^3\theta} -
                       a^2 \sin\theta
                 \right),                                 \\
   \frac{d\phi}{d\sigma}   & = &
      - \left(a - \frac{\xi}{\sin^2\theta}\right) +
      \frac{a}{\Delta}
           \left(r^2 + a^2 - \xi a \right),
                     \label{eq11}
\end{eqnarray}
where $\eta = Q/M^2E^2$ and $\xi = L_z/ME$ are the
Chan\-dra\-sekhar's constants\cite{chandra}, which should be
derived from the initial conditions in the disk plane; $r$ and
$a$ are the appropriate dimensionless radial variable and constant
of rotation respectively. The system (\ref{eq6})-(\ref{eq11}) has
also two integrals,
\begin{eqnarray}
  \epsilon_1 & \equiv & r_1^2 - r^4 -
      \left(a^2 - \xi^2 - \eta\right) r^2 -   % \nonumber \\
 %            &        &
%      \phantom{.} -
      2\left[\left(a - \xi\right)^2 + \eta \right] r +
      a^2\eta = 0,                  \\
  \epsilon_2 & \equiv & \theta_1^2 - \eta - \cos^2\theta
      \left(a^2 - \frac{\xi^2}{\sin^2\theta}\right) = 0,
                    \label{eq13}
\end{eqnarray}
which can be used for the precision control.
This method differs from the approach which was developed  in
papers.\cite{cunn1,cunn2,karas1,rauch}

%\section{Numerical method}

First, one should define
the Chandrasekhar's constants for each quantum and then
integrate the system (\ref{eq6})-(\ref{eq11}) to either
the infinity or the events horizon, depending on the
constants values.
%}}

%\mbox{ \parbox{\columnwidth}{
    We assume that the hot ring emits quanta
which are distributed by isotropic way
in its local frame.
The simulations are based on the trajectory classification,
depending on the Chan\-dra\-se\-khar's constants.\cite{zakh86,zakh89}

%\section{Simulation results}

\begin{figure*}[!t]
\begin{center}
\includegraphics[width=0.98\textwidth]{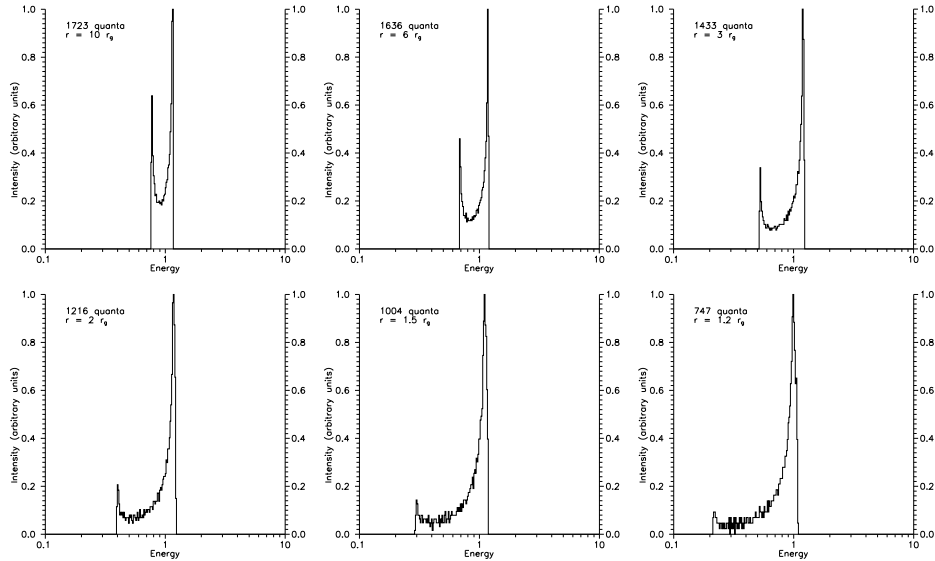}
\end{center}
 % \psfig{figure=specaa09.ps,width=12cm}
%  \vspace{-7mm}
  \caption{Spectrum of the hot ring for $a=0.9$, $\theta=60^\circ$
           and different radial coordinates.
           The marginally stable orbit lays at $r=1.16~r_g$.}
  \label{spectrum09}
\end{figure*}

     The simulated spectrum of a hot ring for $a=0.9$,
$\theta = 60^\circ$ and different radius values is shown in
Fig.~\ref{spectrum09}. The proper quantum energy (in co-moving
frame) is set to unity. The observer at infinity registers the
characteristic two-peak profile, where the ``blue'' peak is
higher than the ``red'' one and the center is shifted to the
left. Some spectrum "oscillations" near its minimum is explained
by pure statistical reasons and has no the physical nature.

     As far as the radius diminishes the spectrum is enhanced, i.e.
increases the residual between the maximum and minimum quanta
energy, registered by a distant observer. For example, for
$a=0.9$, $r=1.2~r_g$ and $\theta=60^\circ$, where $r_g$ has its
standard form $r_g=2kM/c^2$, i.e. in the vicinity of the
marginally stable orbit, the quanta, flown out to the distant
observer, may differ 5 times in their energy. The red maximum
decreases its height with diminishing the radius and at $r<2~r_g$
is almost undistinguished. It is interesting to note that the
spectrum has very sharp edges, both red and blue. Thus, for
$a=0.9$, $r=3~r_g$, $\theta=60^\circ$ the distant observer has
registered 1433 quanta of 20417 emitted by isotropic way; 127 of
them ($\approx$ 9\%) drop to the interval $1.184 < E < 1.202$
(blue maximum) and 43 quanta drop to $0.525 < E < 0.533$ (red
maximum), whereas no one quantum has the energy $E < 0.518$ or $E
> 1.236$.

%\looseness=-1

\begin{figure*}[!t]
\begin{center}
\includegraphics[width=0.98\textwidth]{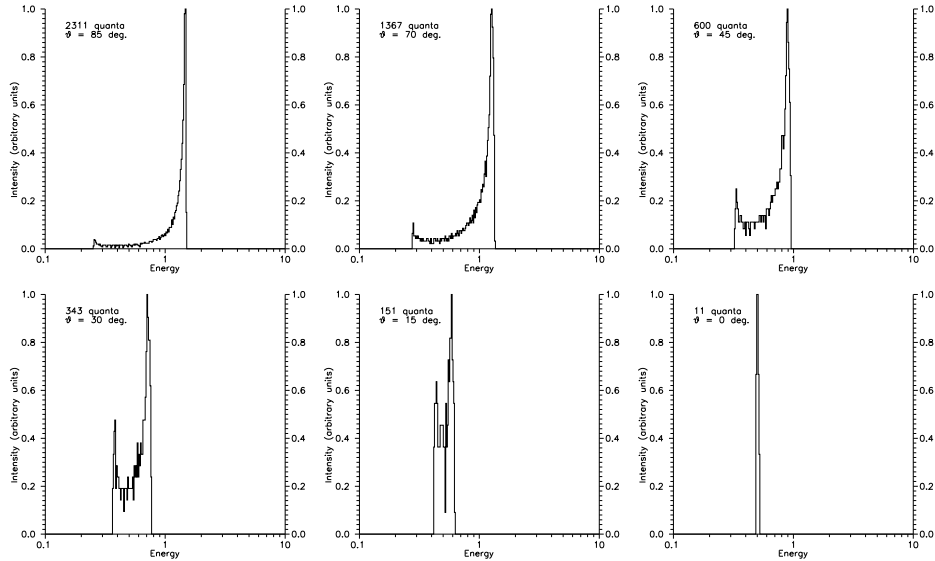}
\end{center}
%  \psfig{figure=specaa15.ps,width=12cm}
 %\vspace{-7mm}
  \caption{Spectrum of a hot ring for $a=0.9$, $r=1.5~r_g$
           and different $\theta$ angles.}
  \label{spectrumtheta}
\end{figure*}

  A spectrum of a hot ring for $a=0.9$, $r=1.5~r_g$ and different
$\theta$ values is shown on Fig.~\ref{spectrumtheta}. The spectrum
for $\theta=60^\circ$ and the same $a$ and $r$ values is included
in Fig.~\ref{spectrum09} and should be added to the current figure
too.  As it follows from the figure, the spectrum critically depends
on the disk inclination angle.
For large $\theta$ values, when the line of sight slips almost along
the disk plane, the spectrum is strongly stretched, its red maximum
is essentially absent, but the blue one appears narrow and very high.
The red wing is strongly stretched because of the Doppler effect,
so that the observer registers the quanta with 5 times energy
difference. As far as the $\theta$ angle diminishes the spectrum
grows narrow and changes the shape: its red maximum first appears
and then gradually increases its height. At $\theta=0^\circ$ both
maxima merge to each other and the spectrum looks like the
$\delta$-function. It is evident because all the points of
the emitting ring are equal in their conditions with respect to the
observer. The frequency of registered quanta in that case is 2 time
lower than the frequency of the emitted ones. A fall in frequency
consists here in two effects, acting in the same direction: the
trans\-versal Doppler effect and the gravitational red shift.

\begin{figure*}[!p]
%  \centerline{
  %\psfig{figure=spec9070.ps,width=14.8cm}
 %            }
\begin{center}
\includegraphics[width=0.98\textwidth]{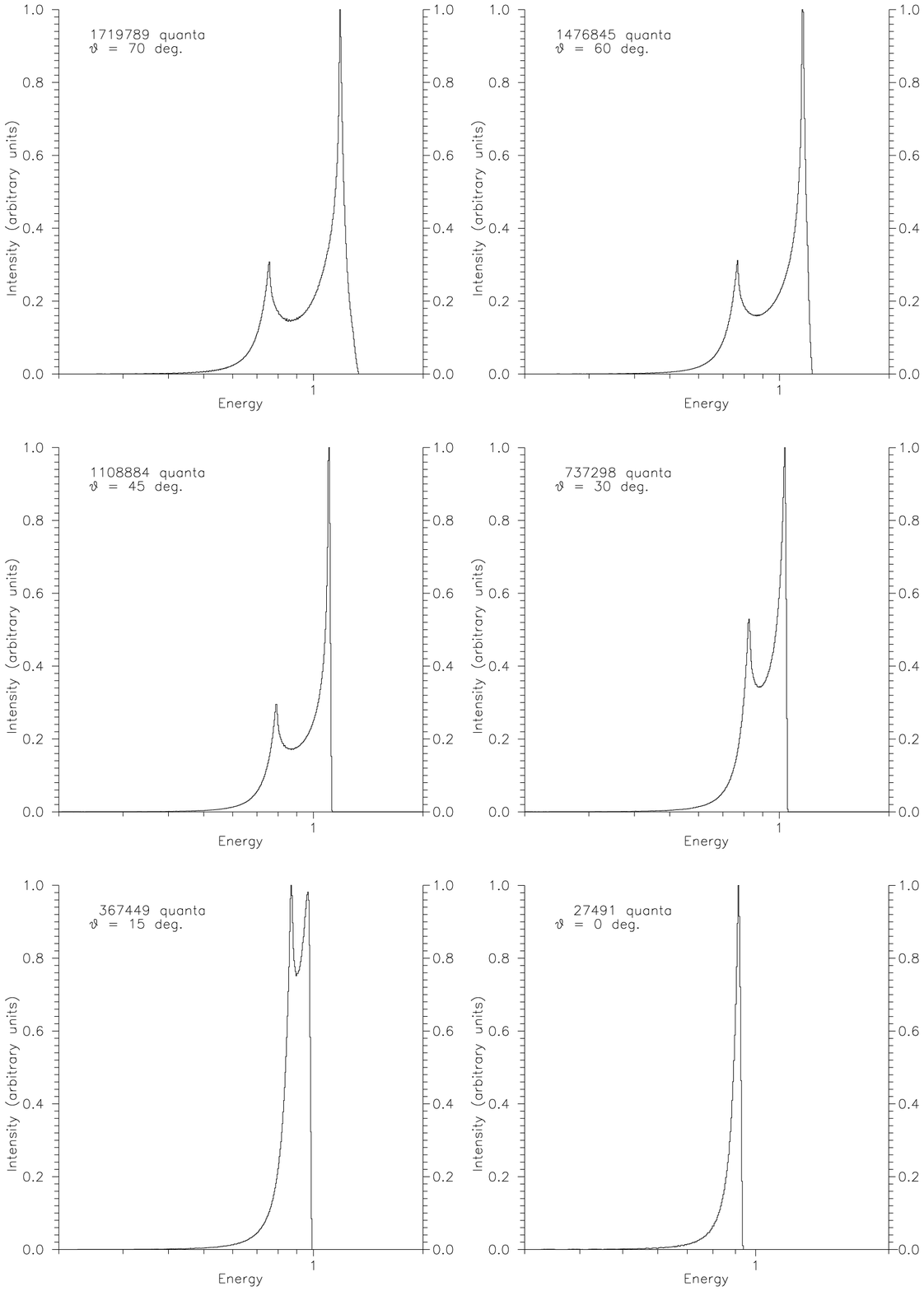}
\end{center}
 \vspace{-7mm}
  \caption{ The spectral line shapes for different  $\theta$ angles.
The emitting region is the wide ring and its inner boundary is the last
stable orbit (for rotational parameter $a=0.9$ this $r$-value is equal
to $r=1.16\,r_g$), its outer boundary corresponds to $r=10\,r_g$.}
  \label{spec9070}
\end{figure*}

\begin{figure*}[!p]
\begin{center}
\includegraphics[width=0.98\textwidth]{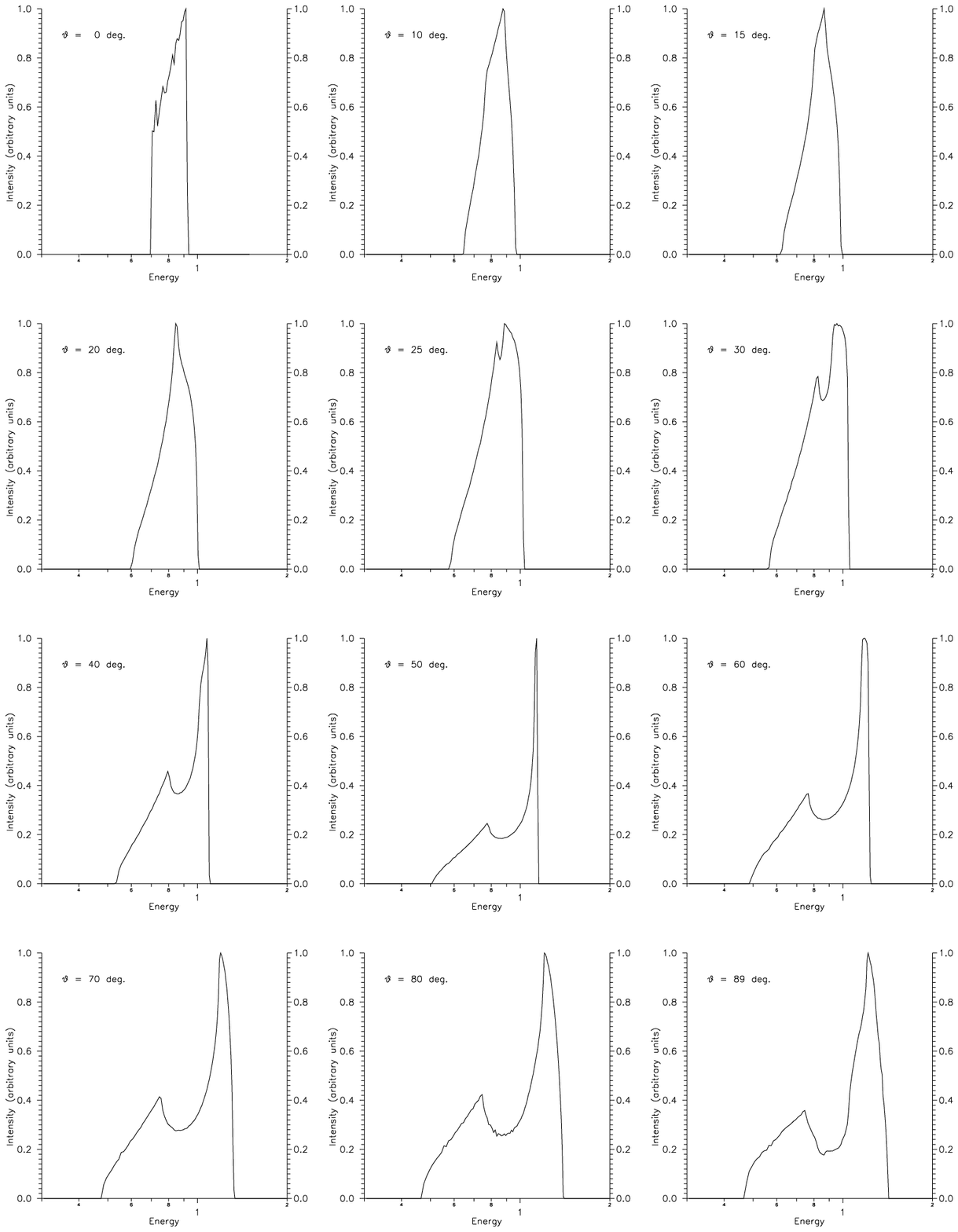}
\end{center}
  \caption{The spectral line shapes corresponding
to an accretion disk
with outer and inner radii ($r_{\rm in} = 3\,r_g$ and
$r_{\rm out} = 10\,r_g$)
in Schwarzschild black hole field
for different position angles of a distant observer.
The temperature is distributed according to the Shakura -- Sunyaev model.}
  \label{spec001}
\end{figure*}

%\section{Discussion and conclusions}

The strong variability of Seyfert galaxies in X-ray  does
not contradict the assumption, that we observe the emission of the
hot rings from the inner region of accretion disk, which can decay or
grow dim, going towards a horizon as time passes. The spectrum dynamics
is understood qualitatively by reference to
Fig.~\ref{spectrum09},
considered sequentially
from top to bottom.

To analyze an influence of a disk width
on the shapes of the line  we consider
the case of a wide accretion disk
and it was shown that the shape of the spectral line retains its
type with two peaks\cite{zak_rep2,zak_rept} (see Fig.\ref{spec9070}).
It is noted that the inner
parts give the essential contribution into red wing of spectrum.

It is known that the standard disk models (like, for example,
Shakura -- Sunyaev and Novikov -- Thorne disk models)
hardly ever could be used to describe temperature distributions
in accretion disks of Seyfert galaxies, however to
show an influence of a temperature distribution
on the spectral line shapes we use the standard disk model
as a template.
Fig. \ref{spec001} demonstrates the shape of emitted monochromatic
line in Schwarzschild black hole field with temperature distributed
according to $\alpha$-disk model.\cite{shasun}

Details of computations and a full list of references could be
found in papers.\cite{zak_rep1,zak_rep2,zak_rep3} An application
of such approach to estimate magnetic fields in AGNs and
microquasars is described in details.\cite{ZKLR02}

\section*{Acknowledgements}

We would like to thank   the Organizers of the Xth ICRA Workshop
and especially prof. V.~Gurzadyan for the kind and warm
hospitality in Rome and Pescara. AFZ would like to thank
Dipartimento di Fisica Universita di Lecce and INFN, Sezione di
Lecce where the final version of the paper was prepared.
 This work was supported in part by Russian Foundation for Basic
Research (project N 00-02-16108).

 \end{document}